# The Structure and Dynamics of Flexible Polyelectrolyte Combs

A.Papagiannopoulos[1], T.A.Waigh[1*], C.Fernyhough[2], T.Hardingham[3], M.Heinrich[4]

[1] Department of Physics and Astronomy, University of Leeds, Leeds, LS2 9JT, UK.

[2] Department of Chemistry, University of Sheffield, Sheffield, Brook Hill, Sheffield, S3 7HF, UK.

[3] School of Biological Sciences, Michael Smith Building, University of Manchester, Oxford Road, Manchester, M13 9PT, UK.

[4] Institut für Festkörperforschung, Forschungszentrum Jülich, D-52425 Jülich, Germany.

(*) To whom correspondence should be addressed. E-mail: t.a.waigh@leeds.ac.uk

The structure and dynamics of a range of polystyrene sulphonate comb polyelectrolytes with well defined chain architectures were examined with static light scattering, dynamic light scattering, small angle neutron/X-ray scattering, particle tracking microrheology and diffusing wave spectroscopy. The chains adopted extended cylindrical conformations in dilute solutions. In semi-dilute solutions a universal behaviour was found for the correlation length ($\xi$) on the monomer concentration (c), $\xi=(3.0\pm0.1/b)c^{-0.47\pm0.01}$, independent of the comb architecture (b is the monomer size). The high frequency viscoelasticity ($10^4$-$10^6$Hz) is found to be in agreement with a model for the Rouse dynamics of the chains and is again independent of the architecture of the combs. However the architecture had a significant impact on the low frequency viscosity of the solutions and the particle tracking data was in good agreement with a dynamic scaling theory for the unentangled dynamics of the combs. Analogous results are found with the biological comb polyelectrolyte aggrecan.

Polyelectrolytes continue to be one of the least well understood areas of soft-condensed matter physics [1]. The electrostatic interaction between monomers is long ranged and complicates the theoretical understanding of the chain morphology and dynamics. A wide range of physical phenomena are associated with the charge interactions along a polymeric backbone including counterion condensation [2], like-charge attraction [3], electrophoresis [4], the Rayleigh charge instability [5] and complexation [6].

The structure and dynamics of comb polyelectrolyte solutions provide many open questions. Scaling theory has been successfully applied to the behaviour of flexible linear polyelectrolytes [7, 8]. The theory has not yet been extended to branched charged morphologies, although progress has been made with their neutral counterparts [9, 10, 11]. Synthetic anionically-polymerised sulphonated polystyrene combs (PSS) with well defined side-chain and main-chain lengths offer an ideal system to extend the physical understanding of branched flexible polyelectrolytes [12].

The corresponding biological comb system, the proteoglycans, remains one of the least well understood biopolymer systems in molecular biology [13, 14, 15] e.g. aggrecan, and the mucins. This is due to a series of complicating factors including chemical heterogeneity, noncrystallinity, and long-range charge interactions.

Previous experiments have examined concentrated solutions of porcine stomach mucin solutions [13]. Evidence was found for liquid crystallinity with these charged biological comb polymers in semi-dilute and concentrated solutions [16]. A rich variety of phase behaviour is thus possible with comb polyelectrolytes.

To date there is only a small amount of information available on the behaviour of solutions of synthetic side-chain liquid-crystalline and isotropic comb polymers. This contrasts with the case of the melt dynamics of comb polymers, which have experienced a number of experimental studies [17, 18, 19] and find complex dynamic phenomena induced by the interplay of the activated reptative motion of the polymeric backbone and the nematic order parameters for the backbone and side chains.

From a practical point of view, we are motivated in the research on comb polyelectrolytes by biomedical questions relating to the functioning of aggrecan in diseased materials [20]. The loss of the charge fraction in aggrecan molecules is a contributing factor in osteoarthritis conditions [21] and the highly viscous shear modulus of aggrecan is vitally important in its action as the compliant matrix between collagen fibres in composite cartilaginous tissues [22]. Aggrecan aggregates are giant polyelectrolyte comb polymers with a hyaluronic acid back-bone and pendant side-chain molecules [20]. The side-chain molecules are themselves comb polymers with protein backbones and charged carbohydrate side-chains [23]. The side-chain combs self-assemble onto the hyaluronic acid backbone through a pH switchable process, which is driven by non-specific electrostatic interactions. A quantitative connection between the molecular structure and the rheological properties of aggrecan has not yet been developed.

In the current study we employed a range of non-invasive experimental techniques to study the structure and viscoelasticity of comb polyelectrolytes in solution. The viscoelasticity was studied at low frequencies using the method of video particle tracking microrheology [24, 25] with 0.5µm carboxylate coated polystyrene spheres. The microrheology data was in agreement with both bulk drag rheometry and two-particle cross-correlation methods [24] validating the measurements. Diffusing wave spectroscopy was used in the transmission geometry [26] with 0.5µm amino functionalised polystyrene spheres to probe the high frequency linear viscoelasticity ($10^4$-$10^6$Hz) of the combs.

The *dilute morphology* of the PSS comb polyelectrolytes was deduced by fitting combined static light scattering and neutron scattering data to a model of cylinders (**figure 1**). Spherical and ellipsoidal models were not compatible with the scattering data and the length (L) and radii (R) of the cylinders scaled qualitatively as expected with the different branch lengths of the five PSS comb architectures; L and R were in the range 2044-3113Å and 249-303Å respectively and the number of side-chains ranged from 9-150. Addition of salt (1M NaCl) to the PSS samples caused a contraction of their conformations (~30%), but they continued to adopt extended cylindrical conformations, indicating the dominant effect of excluded volume interactions on the comb morphologies. Molecular weights from the fits were in agreement with the values from GPC measurements.

For overlapping *semi-dilute solutions* of the combs the correlation length ($\xi$) of the polyelectrolyte solutions was extracted from the position of the peak (q*) in both small angle X-ray scattering (PSS) and small angle neutron scattering (aggrecan) experiments ($\xi=2\pi/q^*$, inset **figure 2**) [27]. The theoretical prediction [7] for the correlation length of semi-dilute solutions of linear flexible polyelectrolytes is

$$\xi \approx b(cb^3)^{-1/2} B^{1/2} \qquad (1)$$

Where $b$ is the monomer length (for PSS, b=3Å) [28], c is the polymer concentration and B is the ratio of the chain's contour length to its size in dilute solution.

The linear and comb data of PSS for the correlation length ($\xi$) as a function of polymer concentration all fall on the same line (**figure 2**), $\xi=(3.0\pm0.1/b)c^{-0.47\pm0.01}$. The peak observed in the SAXS profiles is predominantly due to the conformation of the side-chains, since they have a much higher mass fraction than the main-chains. The correlation length observed with polyelectrolyte combs thus reflects the conformation of the side-chains in semi-dilute solution. From the similarity between the data for different comb architectures and linear PSS [28] we conclude that there is no change in side-chain flexibility or conformation in semi-dilute solutions between the different combs architectures. Aggrecan solutions (**figure 2**) again follow the predicted behaviour for the correlation length, equation (1), and again the scattering is expected to be dominated by the side-chains.

At *high frequencies* diffusing wave spectroscopy experiments show that both the storage (G') and loss (G'') moduli for PSS and aggrecan follow the predictions of the dynamic scaling theory for Rouse dynamics of linear polyelectrolyte chains (**figure 3**) [7].

$$G'(\omega) \approx G''(\omega) \approx (\eta_s k_B T)^{\frac{1}{2}} \left(\frac{B}{b}\right)^{-\frac{3}{4}} c^{\frac{3}{4}} \omega^{\frac{1}{2}} \qquad (2)$$

where kT is the thermal energy, $\eta_s$ is the solvent viscosity and ω is the frequency.

The high frequency shear moduli of the comb polyelectrolytes are found to be independent of the comb architecture for both PSS and aggrecan (figure 3, inset). Reasonable agreement is found with the predictions of equation 2. It is concluded that at sufficiently high frequencies ($10^5$ Hz) the dynamics of subsections of the comb chains are only sensitive to their local environment. This result is in agreement with the SAXS/SANS results on the static correlation length showing the insensitivity to global chain topology.

In sharp contrast the *low frequency dynamics* of the polyelectrolyte combs is dramatically affected by their architecture. Particle tracking results for the PSS comb viscosities are shown in **figure 4**. Linear PSS (with a much higher molecular weight than the backbone lengths we examine) is shown for comparison and has the characteristic dilute, semi-dilute unentangled ($\eta \sim c^{0.5}$) and entangled ($\eta \sim c^{1.5}$) regimes evident [7]. With sparsely branched comb PSS both unentangled and entangled regimes are observed, but with much lower entanglement concentrations than seen previously with the linear chain architecture. With the two most densely grafted PSS combs the viscosity demonstrated entangled behaviour immediately above the overlap concentration (c*) and this is also observed with both the assembled and disassembled aggrecan solutions.

A model for the low frequency viscoelasticity of *unentangled* comb polyelectrolytes was developed and was found to be in good agreement with results from SANS/SAXS results for the extension of the chains (B). From the static results the conformation of the polyelectrolyte combs is assumed a random walk of correlation blobs for both the main-chain and side-chains as indicated from the scattering experiments. The slowest relaxation time of the comb molecules is the relaxation time of the main-chain. The relaxation time of a branched blob of the main-chain is no longer the Zimm time of the correlation blob ($t_x$), since there is a side-chain emerging from it. The relaxation time of such a blob is defined by the relaxation of the attached side-chain, which is the Rouse time of the linear side-chain ($\tau^{side}_{Rouse}$)

$$t^{side}_{Rouse} \approx t_x \left(\frac{n}{g}\right)^2 \qquad (3)$$

Where $n$ is the number of monomers on a side-chain, and $g$ is the number of monomers inside a correlation blob. For blobs of the main-chain with no side-chain attached the relaxation time the Zimm time ($t_x$) is

$$t_x \approx \frac{h_s x^3}{k_B T} \qquad (4)$$

The relaxation time of a blob of the main-chain can be written as an average over all the blobs, i.e. over blobs with no side-chains and blobs with side-chains. The fraction (p) of main-chain blobs with side-chains attached to them is

$$p = \frac{s}{N/g} \qquad (5)$$

Where $s$ is the number of side-chains attached to a comb and $N$ is the number of monomers on the main-chain. The average relaxation time of a blob ($\tau_{blob}$) of the main-chain is now given by

$$t_{blob} \approx t_x + \frac{s}{N/g} t_{Rouse} \qquad (6)$$

The main-chain is a random walk of blobs and the Rouse model can be applied. Therefore the relaxation time ($\tau^{comb}_{Rouse}$) for the main-chain is

$$t_{Rouse}^{comb} \approx t_{blob}\left(\frac{N}{g}\right)^2 \qquad (7)$$

The shear modulus of the solutions will be equal to the number of comb molecules per unit volume multiplied by $k_B T$ [7]. Finally the specific viscosity ($\eta_{sp}$) of the combs is given by the product of the slowest relaxation time ($\tau_{comb}$) multiplied by the shear modulus and is therefore

$$\eta_{sp} \equiv \frac{\eta - \eta_s}{\eta_s} \approx \frac{N^2}{K}\left(\frac{B}{b}\right)^{-\frac{3}{2}}\left(1 + \frac{n^2 s}{N\left(\frac{B}{b}\right)^{\frac{3}{2}}} c^{\frac{1}{2}}\right) c^{\frac{1}{2}} \qquad (8)$$

Where $K$ is the total number of monomers on the comb polymer. This relation simplifies to that for the linear case if one sets $n=0$ or $s=0$ and $N=K$.

Implicit in equation 8 is the notion of two overlap concentration; c* and $c^{brush}$. c* is the concentration at which the backbones overlap (c*≈$B^3 b^{-3} N^{-2}$) and the Zimm modes of the backbone become hydrodynamically screened (Rouse-like). $c^{brush}$ is the side-chain overlap concentration where the Zimm modes of the side-chains also become hydrodynamically screened ($c^{brush}$≈$B^3 b^{-3} n^{-2}$). Equation 8 has only one free fit parameter, the extension of the chains (B) and the values are found to be in good agreement with those measured independently from SAXS/SANS measurements for the lightly branched (s=9-24, n=286-329) PSS combs ie. B is 2.6 from viscosity and B is 3.0 from SAXS (figure 4 inset shows the fit).

We have thus demonstrated it is possible to dramatically modify the low frequency rheological behaviour of polyelectrolyte solutions by the inclusion of side-chains and quantitative predictions can be made for the impact of the chain topology on the unentangled rheology of the comb solutions. In contrast the high frequency viscoelasticity and small angle X-ray/neutron scattering were in agreement with that expected for topologically independent sections of linear flexible polyelectrolyte.

Future studies will concentrate on the development of a scaling theory for the reptative dynamics of entangled polyelectrolyte combs and examine the phase behaviour of lyotropic liquid crystalline combs.

**Acknowledgements**

We thank Ralph Colby, Tanniemola Liverpool, Mike Evans, and Tom McLeish for useful discussions, and the EPSRC for funding AP and TAW. The Jeulich reactor, and the Daresbury synchrotron provided invaluable experimental support.

**Figure 1**.
A typical fit of a cylindrical model for the conformation of the PSS comb polyelectrolytes to combined light scattering and small angle neutron scattering data; absolute intensity is shown as a function of momentum transfer (q). A schematic diagram of the comb polyelectrolyte morphology is shown as an inset, radius (R) and length (L).

**Figure 2.**
The polymer concentration dependence of the mesh size ($\xi$) of semi-dilute solutions calculated from the peak position measured in SAXS for comb11 (squares), 12 (circles), 13 (triangles), 14 (diamonds), 16 (down triangles), linear PSS [28] (filled circles) and in SANS for aggrecan (crossed circles). The straight lines are 1/2 power law fits, equation 1. The SAXS data for comb 12 is indicated as an inset, showing scattered intensity as a function of momentum transfer for 0.044, 0.091, 0.19, 0.38, 0.66 and 0.94 M monomer concentration as the peak position moves from left to right. The arrow indicates increasing polymer concentration.

**Figure 3.**
The complex shear moduli G' (squares) and G'' (circles), as function of frequency from diffusing wave spectroscopy experiments for 10mg/ml aggrecan aggregate (red) and 2 mg/ml PSS 11 (blue). The insert includes the prediction of a Rouse model for the shear moduli as a function of polymer concentration compared with the experimental data, equation 2. The high frequency viscoelasticity ($10^5$Hz) of both aggrecan (squares: monomer black, aggregate red) and PSS (circles: comb11 black, comb 12 red, comb 13 green, comb 14 blue, comb 16 purple) is found to be independent of the comb architecture.

**Figure 4.**
The intrinsic viscosity from particle tracking microrheology of the comb polyelectrolytes as a function of polymer concentration. Symbols correspond to PSS comb11 (sparsely branched topology, black unfilled squares), 16 (densely branched topology, red unfilled squares), linear PSS [28] (black filled rectangles) and for aggrecan (green squares). The continuous lines indicate the scaling for entangled flexible polyelectrolyte viscosity (green, $\eta \sim c^{3/2}$), semi-dilute flexible polyelectrolyte viscosity (blue, $c^{1/2}$) and previous analysed linear polyelectrolyte data (black, $c^{5/4}$, $c^{1/2}$, $c^{3/2}$) The inset shows the fit of equation 8 to the viscosity of a lightly branched PSS comb.

**Figure 1**

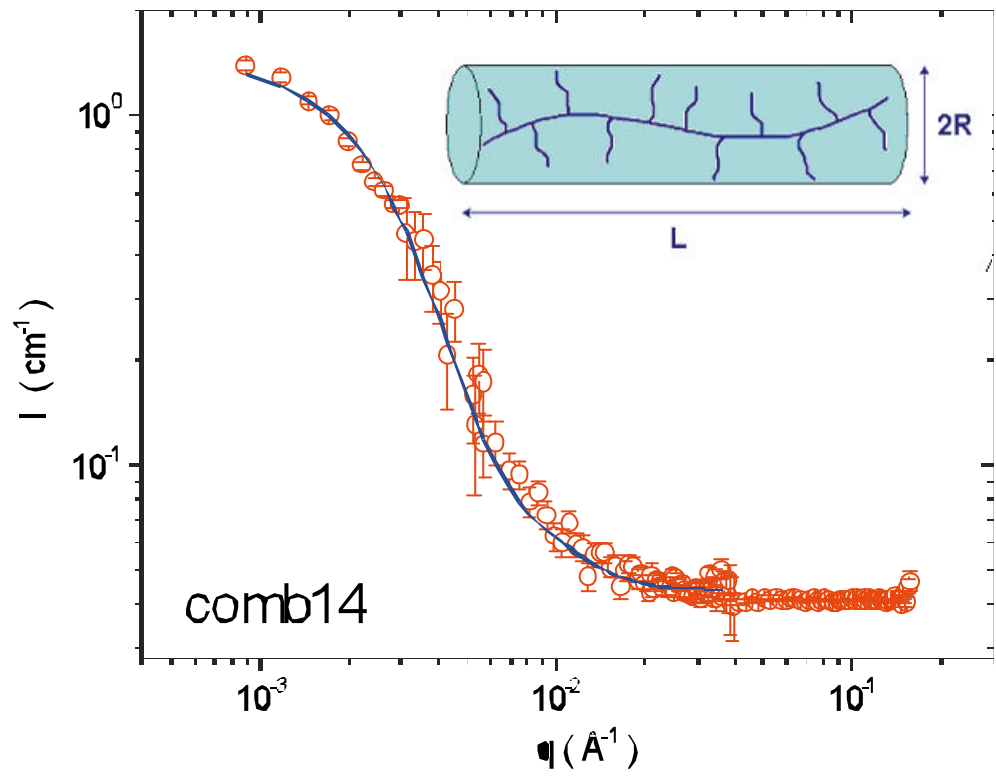

**Figure 2**

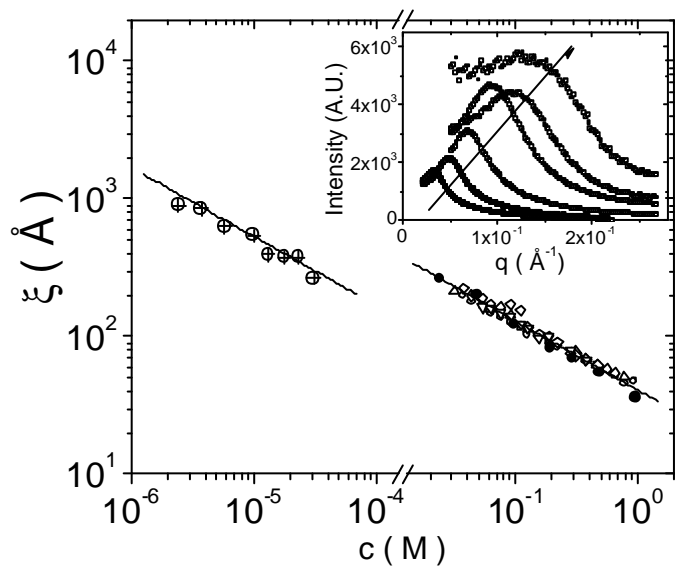

**Figure 3**

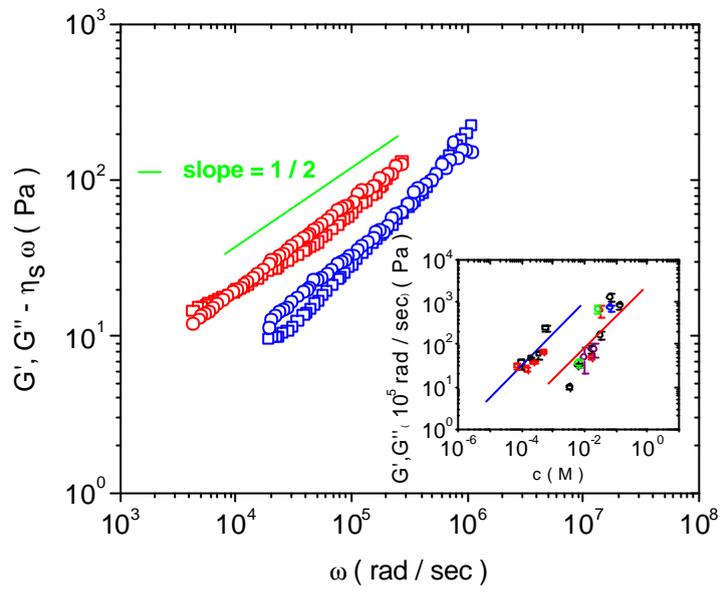

**Figure 4**

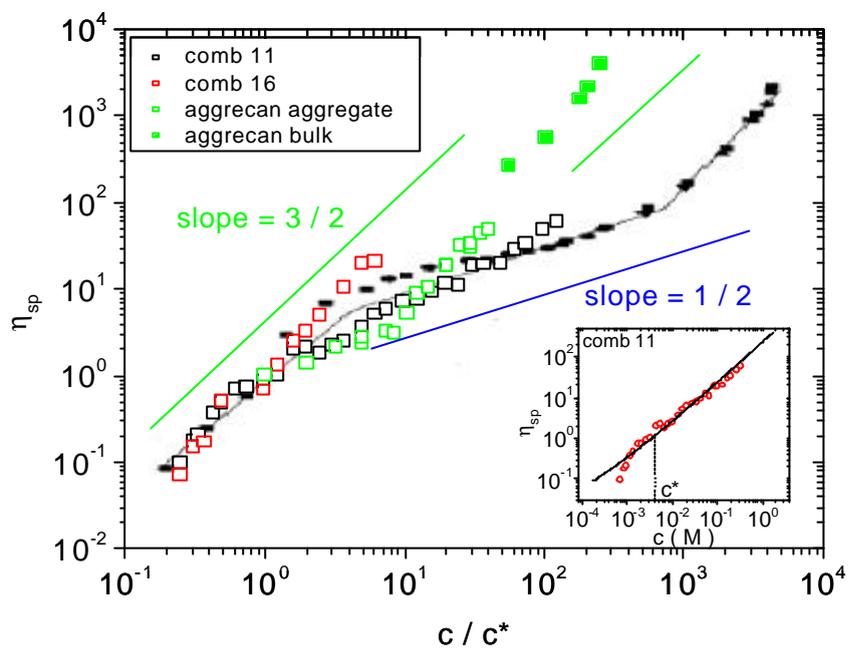